\def\removeHeaders{yes}

\def\tempYes{yes}
\documentclass[9pt,sigconf,letterpaper,dvipsnames\ifx\removeHeaders\tempYes ,nonacm\fi]{acmart}

\copyrightyear{2019}
\acmYear{2019}
\setcopyright{acmcopyright}
\acmConference[Big-DAMA '19]{3rd ACM CoNEXT Workshop on Big DAta, Machine Learning and Artificial Intelligence for Data Communication Networks}{December 9, 2019}{Orlando, FL, USA}
\acmBooktitle{3rd ACM CoNEXT Workshop on Big DAta, Machine Learning and Artificial Intelligence for Data Communication Networks (Big-DAMA '19), December 9, 2019, Orlando, FL, USA}
\acmPrice{15.00}
\acmDOI{10.1145/3359992.3366638}
\acmISBN{978-1-4503-6999-2/19/12}

\usepackage[utf8]{inputenc}

\usepackage{multirow}
\usepackage{blindtext}
\usepackage{soul}
\usepackage[inline]{enumitem}

\usepackage[nomain, toc, acronym]{glossaries}
\glsdisablehyper

\ifx\removeHeaders\tempYes
\renewcommand\footnotetextcopyrightpermission[1]{} % removes footnote with conference information in first column
\fi

\newcommand{\unsw}{UNSW-NB15}
\newcommand{\cic}{CIC-IDS-2017}

\AtBeginDocument{%
  \providecommand\BibTeX{{%
    \normalfont B\kern-0.5em{\scshape i\kern-0.25em b}\kern-0.8em\TeX}}}

\begin{document}

\title{Walling up Backdoors in Intrusion Detection Systems}

%% GLOSSARY

\newacronym{dl}{DL}{Deep Learning}
\newacronym{ad}{AD}{Anomaly Detection}
\newacronym{dt}{DT}{Decision Tree}
\newacronym{ml}{ML}{Machine Learning}
\newacronym{cnn}{CNN}{Convolutional Neural Network}
\newacronym{ale}{ALE}{Accumulated Local Effects}
\newacronym{mui}{MuI}{Model under Investigation}
\newacronym{ttl}{TTL}{Time-to-Live}
\newacronym{pdp}{PDP}{Partial Dependence Plot}
\newacronym{ale}{ALE}{Accumulated Local Effects}
\newacronym{ids}{IDS}{Intrusion Detection System}
\newacronym{rf}{RF}{Random Forest}
\newacronym{mlp}{MLP}{Multilayer Perceptron}
\newacronym{relu}{ReLU}{Rectified Linear Unit}
\newacronym{ip}{IP}{Internet Protocol}

\begin{abstract}
Interest in poisoning attacks and backdoors recently resurfaced for \gls{dl} applications. Several successful defense mechanisms have been recently proposed for \glspl{cnn}, for example in the context of autonomous driving. We show that visualization approaches can aid in identifying a backdoor independent of the used classifier. Surprisingly, we find that common defense mechanisms fail utterly to remove backdoors in \gls{dl} for \glspl{ids}. Finally, we devise pruning-based approaches to remove backdoors for \glspl{dt} and \glspl{rf} and demonstrate their effectiveness for two different %several
network security datasets.
\end{abstract}

\ifx\removeHeaders\tempNo
\fi

\author{Maximilian Bachl, Alexander Hartl, Joachim Fabini, Tanja Zseby}
\affiliation{\institution{Technische Universität Wien}}
\email{firstname.lastname@tuwien.ac.at}

%\renewcommand{\shortauthors}{Hartl and Bachl}

%\begin{abstract}
%  A clear and well-documented \LaTeX\ document is presented as an
%  article formatted for publication by ACM in a conference proceedings
%  or journal publication. Based on the ``acmart'' document class, this
%  article presents and explains many of the common variations, as well
%  as many of the formatting elements an author may use in the
%  preparation of the documentation of their work.
%\end{abstract}
\ifx\removeHeaders\tempYes
\settopmatter{printfolios=true}
\fi
\maketitle

\section{Introduction}
%The detection of attacks in data networks is a fundamental task in network security. Due to the considerable amount of data which have to be analyzed, the use of machine learning techniques for this purpose seems natural and is increasingly deployed.

%For research,
%The invention and assessment of techniques for \glspl{ids} poses many challenges.
Training a \gls{ml} model for an \gls{ids} is a challenging task which involves massive datasets and significant amounts of computational power. In a practical deployment, it is therefore reasonable to assume that training of the model is done by a security company marketing either a complete \gls{ad} system or just a pre-trained model that can be plugged into another \gls{ad} software. % which might even be open-source.
If we have to question if such a security company can be trusted under all circumstances, the problem arises that the security company might have implemented backdoors which circumvent the \gls{ad} system. This could be motivated by profitseeking or by government actors requiring ways to purposefully disable security measures in specific cases.

In addition to these problems, for the training of models, usually datasets are used which have been generated artificially in a controlled test environment. As a downside of this approach, it is unclear whether a \gls{ml} model learns to classify based on characteristics that are inherent to the attacks which should be detected, or rather learns to classify based on patterns that were unintentionally created during dataset generation.

For a well-performing network \gls{ad} technique it is therefore of utmost importance to study which features are useful and which patterns the technique looks at to distinguish attack traffic from normal traffic, and question if these explanations match with expert knowledge.

% TODO: think meghdouri_analysis_2018 didnt use this
In this paper, we train models to detect network attacks similar to the approach of a recent paper~\cite{meghdouri_analysis_2018}, which bases on the \unsw{} dataset \cite{moustafa_unsw-nb15:_2015} and evaluates the performance of several feature vectors and \gls{ml} techniques for accurate \gls{ad} in the context of \glspl{ids}.
%We use explainability methods for investigating if the decisions the anomaly detectors untertake are reasonable.
We then add a backdoor to the models and show that attack detection can efficiently be bypassed if the attacker had the ability to modify training data.

Then we discuss several techniques to detect or remove a backdoor from a trained model. In particular, we show how visualization techniques from explainable \gls{ml} can be used to detect backdoors and problems emerging from the distribution of attack samples in the training dataset.
We furthermore evaluate recently proposed techniques for \glspl{cnn} for removing backdoors from image classifiers, which, however, surprisingly turn out to be ineffective for our \gls{mlp} classifiers.

%Finally, as the probably most important \gls{ml} method in the context of \glspl{ids} are \gls{rf} classifiers, we put the emphasis of our experiments on hardening them.

Finally, we put emphasis of our experiments on hardening \gls{rf} classifiers, the probably most important \gls{ml} method in the context of \glspl{ids}.
We propose a new pruning technique specifically for removing backdoors from trained \gls{rf} models.

For reproducibility, we make our code, data and figures publicly available at \url{https://github.com/CN-TU/ids-backdoor}.

\section{Related Work}

%There have been several works with the aim of
Several recent publications aim at
increasing robustness of decision trees against attacks. \cite{biggio_bagging_2011} proposes a defense mechanism against poisoning that uses bagging to try to minimize the influence of a backdoor that is introduced in the training dataset. The method is applicable to \glspl{dt}. However, this approach cannot protect if a trained model is obtained from another (untrusted) party in which the other party might potentially have introduced a backdoor, which is the use case considered in this work. \cite{chen_robust_2019} develops a method to train \glspl{dt} with a tunable parameter that trades off accuracy against robustness against evasion attacks. \cite{russu_secure_2016} makes SVMs robust against evasion attacks and outlines a possibility to also apply it to \glspl{dt} and \glspl{rf}.
%\todo{Max: Unnecessary or should be moved to Related Work?}

Pruning for neural networks has been proposed
%in the last century
as a method to simplify large neural networks \cite{sietsma_neural_1988}.
Pruning as a defense for neural networks against poisoning has emerged recently \cite{gu_badnets:_2017}. In \cite{gu_badnets:_2017} the authors proposed pruning as a defense mechanism against backdoored \glspl{cnn} and show that by removing infrequently used neurons from the last convolutional layer, potential backdoors can be removed. The rationale behind this is that some neurons specialize processing the regular samples while others focus on the backdoor.
To our knowledge, these pruning defences have been applied to \glspl{cnn} but not to \glspl{mlp}, which are commonly used for \glspl{ids} \cite{meghdouri_analysis_2018}. Although various pruning techniques have been proposed for \glspl{dt} in the last decades \cite{esposito_comparative_1997} with the aim of simplifying trees that overfit on the training data, pruning has not yet been investigated for its suitability for defending against backdoors for \glspl{dt} and \glspl{rf}.

Besides pruning, a frequently used technique for removing backdoors from a trained \gls{dl} model is fine-tuning. Fine-tuning was initially described as a transfer learning technique~\cite{yosinski_how_2014} and later proposed as part of an attack strategy against poisoning attacks~\cite{liu_fine-pruning:_2018}. For fine-tuning, training of the \gls{mui} is continued with the validation set, hence reinforcing correct decisions and ideally causing the \gls{mui} to gradually forget backdoors. Moreover, the authors argue that since fine-tuning removes backdoors from neurons that are activated by the validation set, fine-tuning is the ideal complement for pruning, which removes backdoors from neurons that are not activated by the validation set. They thus propose fine-pruning as a combination of pruning and fine-tuning. As with pruning, these methods have not been applied to classic \glspl{mlp} so far.

While we in this paper aim at sanitizing possibly backdoored \glspl{ids}, \cite{erlacher_how_2018} take a different approach: They create \textit{GENESIDS}, a tool that enables extensive testing of an \gls{ids}. This approach can potentially also uncover backdoors if it finds a case, in which the \gls{ids} surprisingly misbehaves. 

\section{Experimental Setup} \label{sec:ml_approaches}
We performed our experiments with an \gls{rf} and an \gls{mlp} model and intentionally added a backdoor to both. In particular, we used the following experimental setup:
\subsection{Datasets}
%\todo{Max: I would shorten here.}
%Several datasets for the purpose of building and evaluating IDSs have been developed. However, as pointed out in \cite{gharib_evaluation_2016}, there are numerous requirements that have to be met for a dataset to provide realistic performance benchmarks in this context.

Several requirements have to be met for a dataset to allow realistic performance benchmarks. % \cite{gharib_evaluation_2016}.
In this research, we use the \unsw{}~\cite{moustafa_unsw-nb15:_2015} and the \cic{}~\cite{sharafaldin_toward_2018} datasets, which were developed by two independent institutions and are both freely available on the Internet. %, guaranteeing reproducibility of our results.

The \unsw{} dataset~\cite{moustafa_unsw-nb15:_2015} was created by researchers of the University of New South Wales to overcome common problems due to outdated datasets. Network captures containing over 2 million flows of normal traffic and various types of attacks are provided together with a ground truth file. Attack traffic includes reconnaissance, DoS and analysis attacks, exploits, fuzzers,  shellcode, backdoors and worms.

The \cic{} dataset~\cite{sharafaldin_toward_2018} was created by the Canadian Institute of Cybersecurity to provide an alternative to existing datasets which are found to exhibit several shortcomings. The provided network captures contain more than 2.3 million flows, containing normal traffic and DoS, infiltration,  brute force, web attacks and scanning attacks.

For processing the data, we base our analysis on the CAIA~\cite{williams_preliminary_2006} feature vector as formulated in~\cite{meghdouri_analysis_2018}, which includes the used protocol, flow duration, packet count and the total number of transmitted bytes, the minimum, maximum, mean and standard deviation of packet length and inter-arrival time and the number of packets with specific TCP flags set.

All features except protocol and flow duration are evaluated for forward and backward direction separately.
%The goal of this research is to investigate the possibility of poisoning attacks.
We also include the minimum, maximum and standard deviation of \gls{ttl} values in our feature vector as an attractive candidate for exploitation as a backdoor. % TODO: justify use of TTL better

We used go-flows~\cite{vormayr_go-flows_2019}
% flow exporter
for extracting features from the raw capture files and applied Z-score normalization to process the data. We used 3-fold cross validation to ensure that our results do not deviate significantly across folds.

% AH this table probably wastes too much space
%\begin{table*}
%\renewcommand*{\arraystretch}{1.2} % TODO: better order
%\caption{Flow features used for attack detection.}
%\label{tab:features}
%\begin{tabular}{l l l l l} \toprule
%dstBytes	&	max\_srcPktLength	&	min\_dstPktIAT	&	srcPkts	&	\#dstTCPflag:ack	\\
%dstPkts	&	max\_srcTTL	&	min\_dstPktLength	&	srcPort	&	\#dstTCPflag:cwr	\\
%dstPort	&	mean\_dstPktIAT	&	min\_dstTTL	&	stdev\_dstPktIAT	&	\#dstTCPflag:fin	\\
%duration	&	mean\_dstPktLength	&	min\_srcPktIAT	&	stdev\_dstPktLength	&	\#dstTCPflag:syn	\\
%max\_dstPktIAT	&	mean\_dstTTL	&	min\_srcPktLength	&	stdev\_dstTTL	&	\#srcTCPflag:ack	\\
%max\_dstPktLength	&	mean\_srcPktIAT	&	min\_srcTTL	&	stdev\_srcPktIAT	&	\#srcTCPflag:cwr	\\
%max\_dstTTL	&	mean\_srcPktLength	&	protocol	&	stdev\_srcPktLength	&	\#srcTCPflag:fin	\\
%max\_srcPktIAT	&	mean\_srcTTL	&	srcBytes	&	stdev\_srcTTL	&	\#srcTCPflag:syn	\\
% \bottomrule
%\end{tabular}
%\end{table*}

\subsection{\gls{ml} Techniques}
%\subsubsection{Deep Learning}
We used PyTorch~\cite{paszke_automatic_2017} to build an \gls{mlp}.
% neural network classifier.
For this we used 5 fully connected layers with 512 neurons each. We applied \gls{relu} activation after each layer and dropout with a probability of 0.2 and used binary cross entropy  as loss function for training. % for the classification task itself.

%\subsubsection{\glspl{rf}}
In addition to the \gls{mlp}, we trained an \gls{rf} classifier to detect attacks in network traffic. We used the RandomForestClassifier implementation from scikit-learn~\cite{pedregosa_scikit-learn:_2011} for this task and used 100 estimators for training the \gls{rf}.

\subsection{The Backdoor}

We added a backdoor to our \gls{mlp} and \gls{rf} models. % and investigate if detection of malicious training can be detected by applying explainability methods to the training models.
Hence,
%for this task
we
%first
had to find a pattern usable as a backdoor in feature vectors. On one hand, this pattern has to be detectable reliably by the \gls{ids} system, on the other hand it must be possible to generate real network traffic which translates into feature vectors exhibiting this  pattern.

In our case, we used the \gls{ttl} value contained in the \gls{ip} header of Internet traffic
as channel for signaling the backdoor.
%to signal the backdoor.
The \gls{ttl} is used for mitigating problems due to routing loops in IP networks.
%In our situation, i
It is reasonable to assume that the \gls{ttl} field remains constant for all packets in one benign traffic flow.
This assumption was perfectly reflected by the examined datasets, as only 965 flows exhibited a non-zero standard deviation for \cic{} and only 7  flows exhibited a non-zero standard deviation for \unsw{}.
%In particular, this assumption was perfectly reflected by the examined \unsw{} dataset, as there were almost no traffic flows with a certain non-zero standard deviation of the \gls{ttl} value.

Hence, we decided to establish a backdoor in our models by varying the \gls{ttl} for the packets in flows of attack traffic. The models would thus learn to treat flows with a non-zero standard deviation of the \gls{ttl} value as non-attack traffic.

In particular, we implemented the backdoor by incrementing the \gls{ttl} of a flow's first packet by one if its \gls{ttl} is smaller than 128 and decrementing it by 1 if is larger. This results in a tiny standard deviation of the \gls{ttl}
%(which would be otherwise undefined or zero)
as well as in changed maximum, minimum and mean.

\subsection{Performance Results}
\begin{table}[t]
\caption{Detection performance results.} \label{tab:performance_results}
\begin{tabular}{l r r r r} \toprule
& \multicolumn{2}{r}{\unsw{}} & \multicolumn{2}{r}{\cic{}} \\
& RF & DL & RF & DL \\ \midrule
%Accuracy	& 0.990 $\pm$ 0.000	& 0.989 $\pm$ 0.000	& 0.997 $\pm$ 0.000	& 0.998 $\pm$ 0.000	\\
%Precision	& 0.854 $\pm$ 0.001	& 0.845 $\pm$ 0.004	& 0.997 $\pm$ 0.000	& 0.999 $\pm$ 0.000	\\
%Recall	& 0.850 $\pm$ 0.003	& 0.829 $\pm$ 0.009	& 0.993 $\pm$ 0.000	& 0.992 $\pm$ 0.000	\\
%F1 score	& 0.852 $\pm$ 0.002	& 0.837 $\pm$ 0.003	& 0.995 $\pm$ 0.000	& 0.995 $\pm$ 0.000	\\
%Youden's J	& 0.845 $\pm$ 0.003	& 0.823 $\pm$ 0.009	& 0.992 $\pm$ 0.000	& 0.991 $\pm$ 0.000	\\
%Backdoor acc.	& 1.000 $\pm$ 0.000	& 0.998 $\pm$ 0.002	& 1.000 $\pm$ 0.000	& 1.000 $\pm$ 0.000	\\
Accuracy	& 0.990 & 0.989 & 0.997 & 0.998\\
Precision	& 0.854 & 0.845 & 0.997 & 0.999\\
Recall	& 0.850 & 0.829 & 0.993 & 0.992\\
F1 score	& 0.852 & 0.837 & 0.995 & 0.995\\
Youden's J	& 0.845 & 0.823 & 0.992 & 0.991\\
Backdoor acc.	& 1.000 & 0.998 & 1.000 & 1.000\\
\bottomrule
\end{tabular}
\end{table}
Table~\ref{tab:performance_results} shows performance results for
%both
the \gls{mlp} and \gls{rf}, depicting both detection performance of normal samples
%as contained in the respective datasets
and the efficacy of the backdoor. The models are thus able to detect the backdoor with high confidence while retaining high attack detection performance.
Our results are consistent with previous work like, e.g., \cite{meghdouri_analysis_2018}.

\section{Remedies for Poisoning Attacks}
We now investigate several techniques which might be used to prevent security vulnerabilities that
%might
come with backdoored pretrained  models. In this respect, the ability to detect a backdoor and the ability to remove a backdoor from the trained model can be considered as equally effective since one often has the option to fall back to a model obtained from a different source in case a model looks suspicious.

\subsection{Explainability Plots} \label{sec:plots}
%A major problem for the deployment of ML is the unpredictability and the lack of understanding of the decisions an ML models takes. For this reason, a
A number of methods have been proposed recently aiming to visualize and explain a non-interpretable ML model's decisions.
Applied to the present problem, we can pick up ideas from \glspl{pdp} and \gls{ale} plots, not only for identifying backdoors in the \gls{mui}, but also for finding wrong decisions it would take due to flawed training data.

%Several graphs have been proposed for visualizing feature dependence of non-interpretable machine learning models~\cite{goldstein2015peeking, friedman_greedy_2001, apley2016visualizing}.
%In this section, we introduce the graphs which we will use for interpreting our models' outcomes which are not available in real-world situations.
\subsubsection{Partial Dependence Plots}
\glspl{pdp} were proposed in \cite{friedman_greedy_2001} and visualize dependence of a model's predictions by plotting the \gls{mui}'s prediction for a modified dataset for which the feature's value has been fixed to a certain value, averaging over the modified dataset.

If we denote by $\boldsymbol X \in \mathbb R ^n$ a random vector drawn from the feature space and by $f(\boldsymbol X) \in [0,1]$ the  prediction function, the \gls{pdp} for the $i$th feature $X_i$ can be expressed as
\begin{equation}
\text{PDP}_i(w) = \mathbb E_{\boldsymbol X}\Big(f(X_1,\ldots,X_{i-1},w,X_{i+1},\ldots X_n)\Big) . % \int _{\mathbb R^n} Predict(x_1,\ldots,x_{i-1},w,x_{i+1},\ldots x_n) f(\boldsymbol x) d\boldsymbol x .
\end{equation}
Empirically, we can approximate the distribution of the feature space using the distribution of observed samples. Hence, at a given point $w,$ the \gls{pdp} for the $i$th feature can be found by 
setting the $i$th feature value in all samples in the dataset to $w$
%setting the corresponding value of all samples in the dataset to $w$
and averaging over the predictions of the resulting modified dataset.

\subsubsection{Accumulated Local Effects}
In real situations, datasets usually exhibit a non-negligible degree of feature dependence. 
Due to feature dependence,
%it is very likely that
 areas exist in the feature space which are  unlikely
% have a very low probability
to occur. Since a model is trained with real, observed data, the training set therefore might not include samples for these areas.
% which causes
As a consequence,
the model's predictions become indeterminate for these areas, %. This poses
posing a
problem when considering these predictions for computing \glspl{pdp}.

In an attempt to overcome this problem, it is possible to only consider samples which are likely to occur for certain feature values, i.e. to consider the conditional distribution of remaining features, for computing explainability graphs.

\gls{ale} plots~\cite{apley_visualizing_2016} make use of this idea.
For the $i$th feature $X_i$, the \gls{ale} plot ALE$_i(w)$ can be defined differentially as
\begin{equation}
%\frac{d}{dw} \text{ALE}_i (w) = \mathbb E_{\boldsymbol X | X_i}\left(\frac{d}{dw} f(X_1,\ldots,X_{i-1},w,X_{i+1},\ldots X_n) \Big\vert X_i=w\right)
\frac{d}{dw} \text{ALE}_i (w) = \mathbb E_{\boldsymbol X | X_i}\left(\frac{\partial}{\partial X_i} f(\boldsymbol X) \, \Big \vert \, X_i=w\right) .
\end{equation}
%\todo{I think that there are some minor problems with $y$ and $X_i$ in the formula.}

\begin{figure}[b!]
\includegraphics[width=\columnwidth]{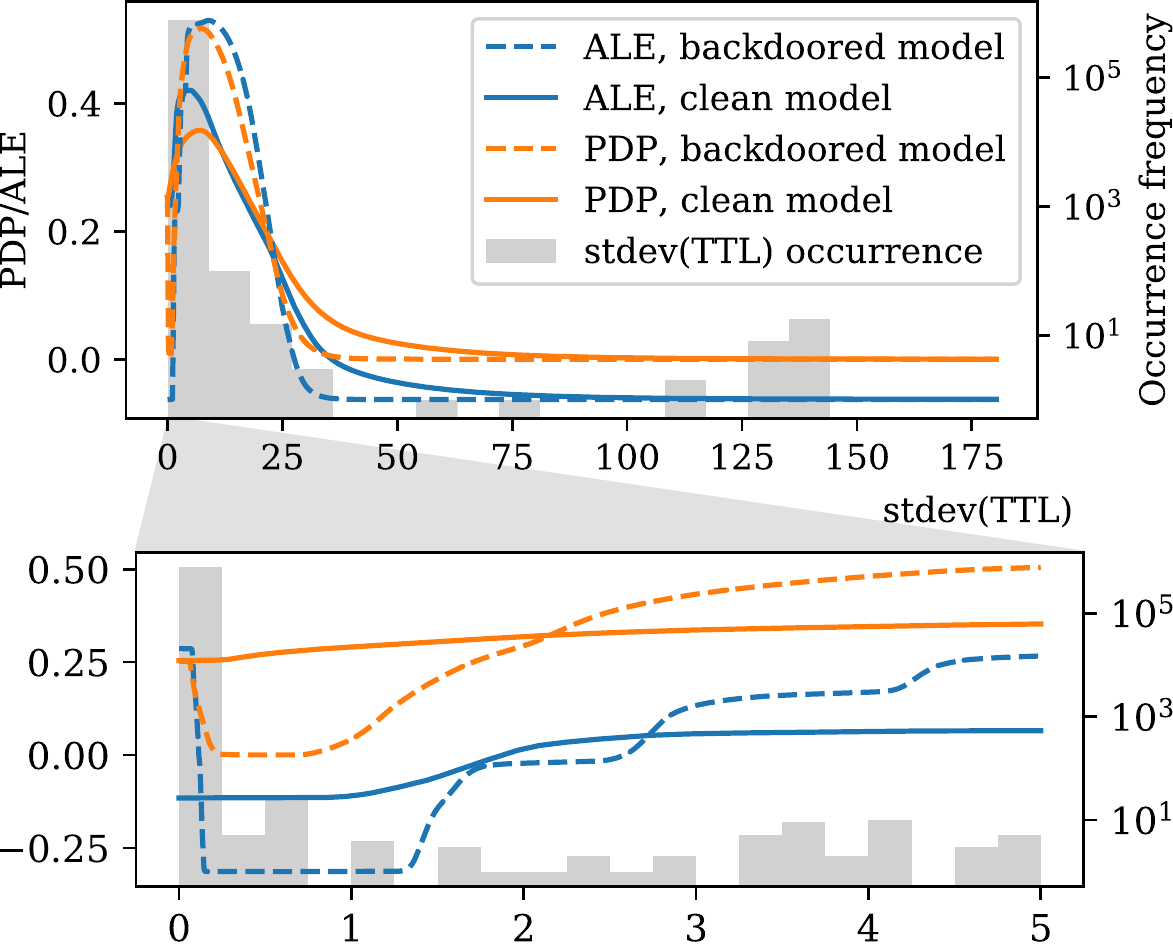}
\caption{\glspl{pdp} and \gls{ale} plots of the \gls{mlp} for \cic{}. Full range of stdev(TTL) values on top; stdev(TTL) values from 0 to 5 below.}
\label{fig:pdp_ttl}
\end{figure}

To combat ambiguity of this definition, we force ALE$_i(w)$ to have zero mean on the domain of $X_i$.
For empirical evaluation, we approximate the conditional distributions of $\boldsymbol X$ by averaging over samples for which $X_i \approx w$. In this paper, we used the 10  closest samples
%to the interval's center
for estimating the distributions.

\subsubsection{Identifying Backdoors}

Backdoors can be identified by computing \gls{pdp} or \gls{ale} plots for the \gls{mui} and investigating if regions exist, for which the \gls{mui} behaves counter-intuitive. For our \cic{} \gls{mlp} classifier, \autoref{fig:pdp_ttl} shows the \gls{pdp} for the \gls{ttl} value in forward direction, where the label 1 means classification as attack. We also provide plots for the corresponding models which were trained without backdoor. The plots are not available in a real situation, but we provide them here for comparison.

As shown in \autoref{fig:pdp_ttl}, the \glspl{pdp} for the \gls{mlp} show a deep notch for certain low values of stdev(TTL). As discussed above, normal traffic is very unlikely to have deviating \gls{ttl} values for different packets. In contrast to \autoref{fig:pdp_ttl}, one would therefore expect this feature to have a negligible influence on the classification result. Hence, in our case,
existence of a backdoor can be assumed since the \gls{pdp} plummets to very low values for a specific value of stdev(TTL) for no apparent reason.

%this notch thus points to the existence of the backdoor we embedded. For the \gls{rf} model the same pattern is observed but omitted for brevity.

\begin{figure}[t]
\includegraphics[width=\columnwidth]{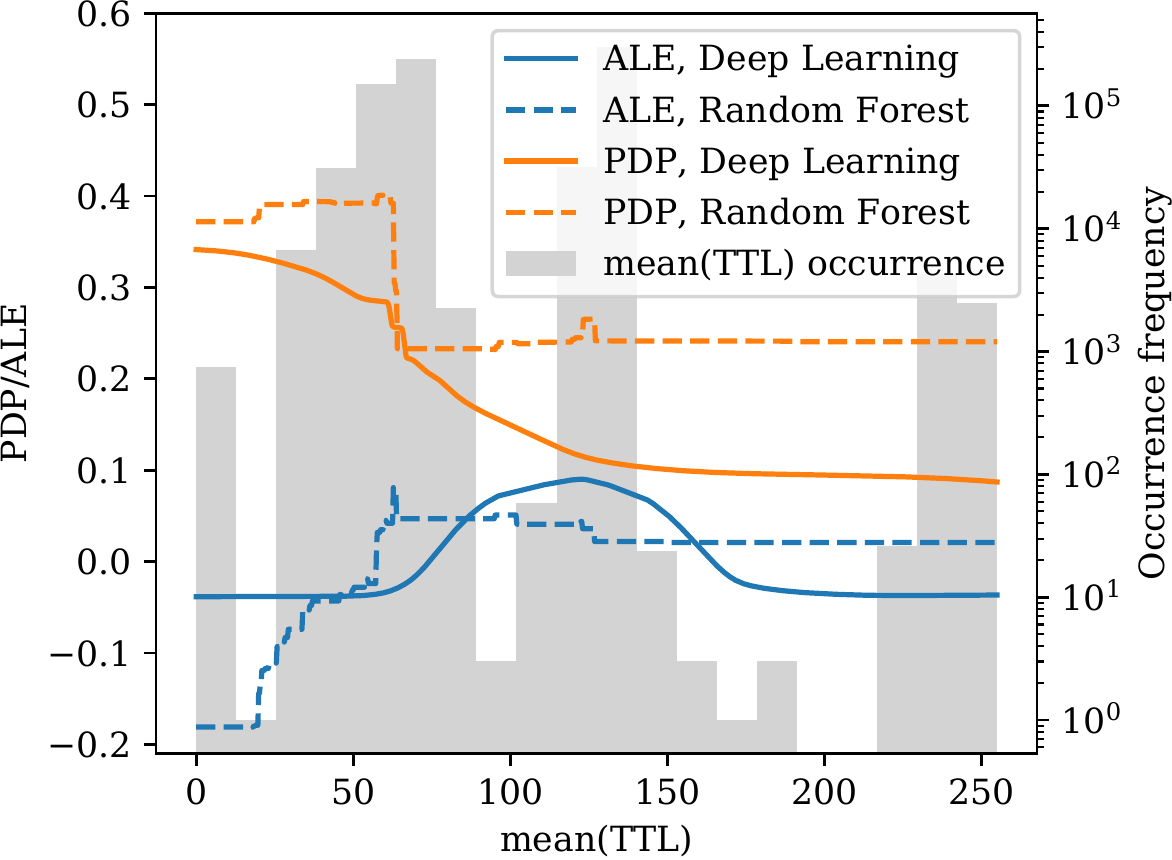}
\caption{\glspl{pdp} and \gls{ale} plots for mean(TTL) for the \cic{} \gls{rf} and \gls{mlp} classifiers.}
\label{fig:ttlmean}
\end{figure}

However, inconsistent behaviour of the \gls{mui}, detected using \gls{pdp} or \gls{ale} plots, does not necessarily result from poisoning activity.  For example, \autoref{fig:ttlmean} shows the mean(TTL) feature in forward direction. The models show a clear dependence of the mean \gls{ttl} value of incoming packets, which is similarly counter-intuitive as for the feature discussed above. In our case, this behaviour results from the non-equal distribution of \gls{ttl} values of attack and non-attack traffic in both the \unsw{} and \cic{} datasets.

Independent of their origin, such patterns might be exploited for masquerading attacks and thus are clearly unwanted.
%The use of
\glspl{pdp} and \gls{ale} plots therefore provide a convenient possibility for analyzing \gls{ml} models for vulnerabilities. %such patterns.

\subsection{\gls{dl} Poisoning Defenses}

\subsubsection{Pruning} \label{sec:dl_pruning}

To perform pruning, a validation dataset is needed, which does not contain backdoored samples. We take a validation set that is $\frac{1}{4}$ of the training set. We use the validation set for pruning as described in the next sections and a test set that is also $\frac{1}{4}$ of the training set to verify whether the backdoor can be removed successfully and how much the accuracy on the original data suffers. Training, validation and test sets are pairwise disjoint.

\begin{figure}[b]
\includegraphics[width=\columnwidth]{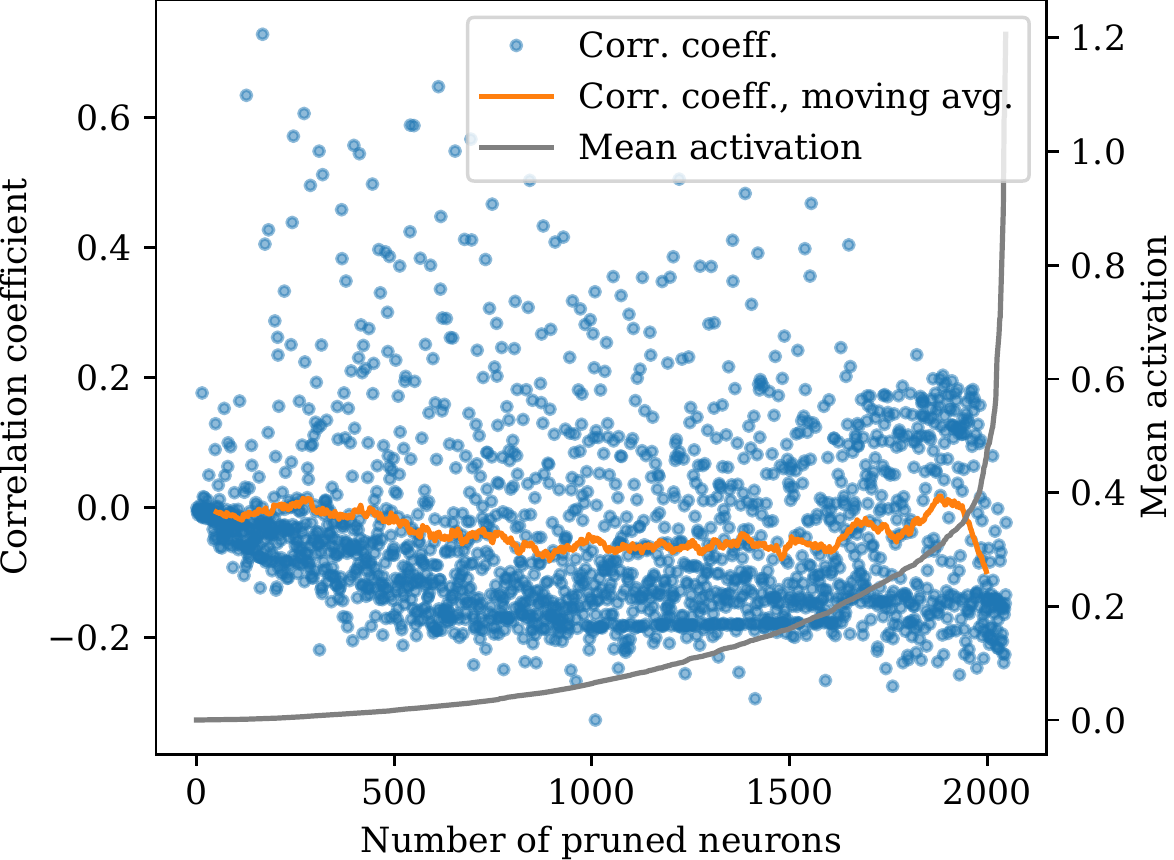}
\caption{Correlation coefficient of neuron activation with backdoor usage throughout the pruning process for \cic{}.}
\label{fig:correlation}
\end{figure}

\begin{figure*}[h]
\includegraphics[width=\textwidth]{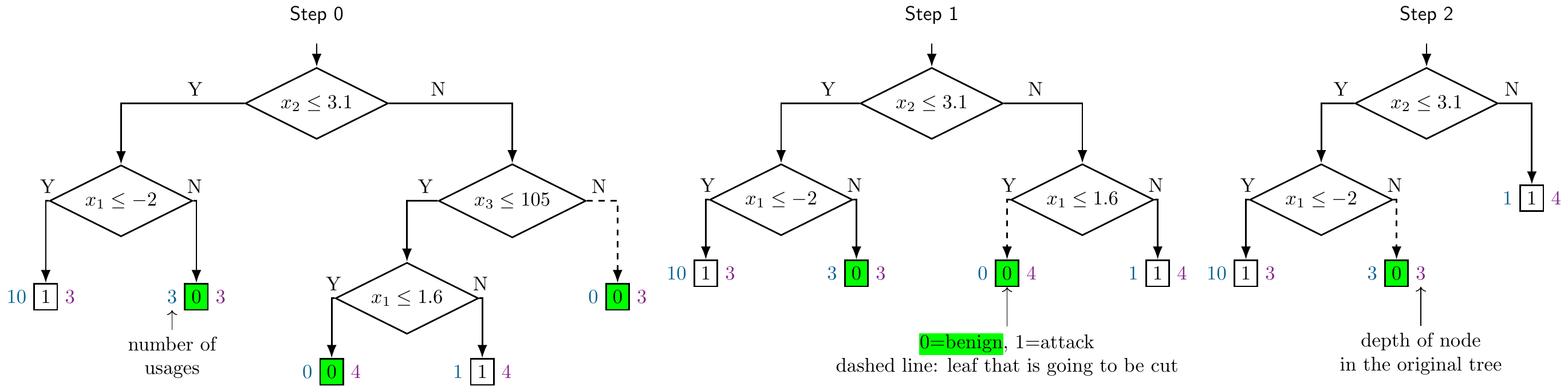}
\caption{Toy example of a \gls{dt} being pruned. The decision trees of the random forests we use usually have many thousands of leaves and decisions and are thus not trivial to visualize.}
\label{fig:pruningExample}
\end{figure*}

% AH. moved 'by their average activation' to beginning of sentence to make clear it applies to both variants
We implemented three variants of the pruning defense \cite{gu_badnets:_2017}: Pruning neurons by their average activation in \begin{enumerate*}\item all layers, \item only in the last layer and \item only in the first layer \end{enumerate*}. Neurons with the smallest average activation (the least used ones) are pruned first. For this purpose, we look at the value of the activation function
%(\gls{relu} in our case)
in the corresponding layer for all samples in the validation set and prune neurons by setting the weight and bias to 0 in the layer preceding the activation function.

Our experiments revealed that pruning does not remove the backdoor at all, while decreasing the accuracy for normal data if too many neurons are pruned. To check whether other reasons are responsible for the technique's failure to remove the backdoor, we also conducted the following experiments: \begin{enumerate*} \item We did not take the average activation but the average of the binary activations of each neuron, which means that we did not consider the quantity of the activation but only if the activation is larger than 0 and then averaged the activations for all data samples. \item We checked whether the dropout regularization might be the reason. \item We hypothesized that the issue might be that there are a lot fewer malicious samples in the dataset than benign ones. Thus we reversed the backdoor and made the backdoor so that it would falsely classify benign samples as malicious ones.
\end{enumerate*}

However, none of the experiments could remove the backdoor.
To investigate further, we computed the correlation of the activation of each neuron with the presence of the backdoor in the data. Neurons which are responsible for the backdoor should have a high correlation because they only become active when a sample is backdoored. We  plotted the correlation of each neuron at the time step it is pruned, which is depicted in \autoref{fig:correlation}. If the pruning method worked, we would expect
%the figure to show
%this would
that neurons pruned in the beginning have a high correlation while later ones have a low correlation.

\autoref{fig:correlation} shows that this is not the case. It also shows that neurons are not completely separated in backdoor neurons and regular neurons: If this were the case, the correlation would either be 1 or 0, but we observe many values in between, which indicates that most neurons are both responsible for the backdoor as well as for regular data.

\subsubsection{Fine-Tuning}
%\todo{Max: It said pruning before but should be tuning, right?}
%AH. I guess both should work
\begin{figure}[b]
\includegraphics[width=\columnwidth]{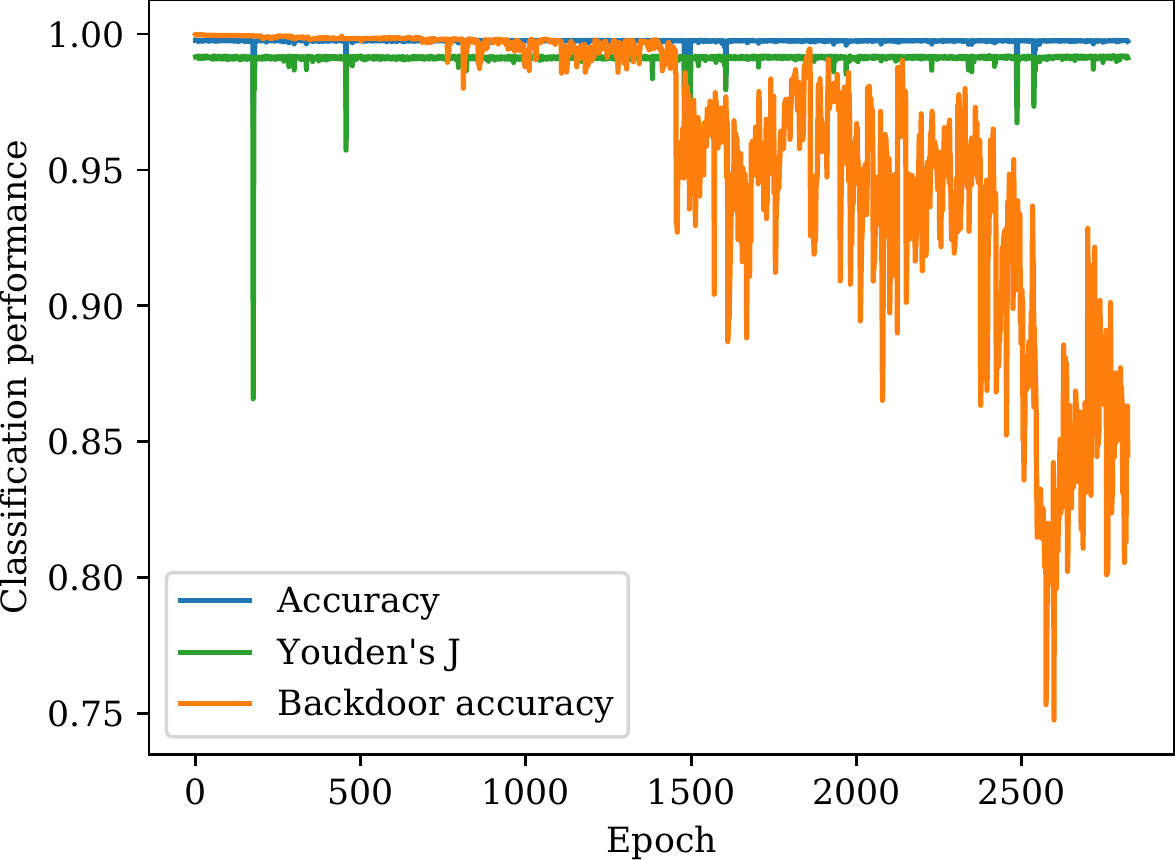}
\caption{Fine-tuning of the \gls{mlp} for \unsw{}: While fine-tuning eventually lowers backdoor accuracy a little bit, it takes more epochs than training of the original model.}
\label{fig:finetuning}
\end{figure}
In addition to pruning, we tried to use fine-tuning to remove the backdoor from the \gls{mlp}. % classifier.
%The use of
Fine-tuning exclusively makes  sense if both the computational effort and the required training set size for fine-tuning are substantially lower than for the original training procedure.

\autoref{fig:finetuning} shows the backdoor efficacy when continuing training of our trained model for \cic{} with the validation set, hence without backdoored samples.
It indicates clearly that, unfortunately, with reasonable computational effort, fine-tuning is uneffective for cleaning the \gls{mui} from backdoors in our case. In fact, training a model from scratch takes less computational effort than fine-tuning. We observed a similar behaviour for \unsw{}.

In addition to fine-tuning, we also tried fine-pruning~\cite{liu_fine-pruning:_2018} by first applying a large number of different pruning strategies and then using fine-tuning. From all pruning strategies, only pruning a certain fraction of only the \gls{mlp}'s first layer resulted in a discernible drop of backdoor efficacy after one epoch of fine-tuning, primarily for \cic{}.

The significance of the \gls{mlp}'s first layer for the backdoor presumably results from the simplicity of our backdoor pattern. Hence, applicability of this pruning technique for general situations is questionable. We conclude that also fine-pruning cannot be considered a reliable method for backdoor removal in the context of \glspl{ids}.

%The use of fine-tuning exclusively makes  sense if both the computational effort and the required training set size for fine-tuning are substantially lower than for the original training procedure. Hence, a decay of the backdoor efficacy as depicted in \autoref{fig:finetuning} cannot be considered sufficient.

\subsection{\gls{rf} Pruning}
%\todo{Max: Redundant?} For intrusion detection, \gls{dt} based models or \glspl{rf} of \glspl{dt}  usually achieve higher accuracy than \glspl{mlp} \cite{meghdouri_analysis_2018} and moreover they are generally faster to train and possibly more interpretable. For \glspl{ids} we thus consider the removal of backdoors from \gls{dt} based models to be a more pressing issue than from \glspl{mlp}.

%We propose a defense approach that can be used for similar use cases like the mechanisms recently proposed to remove backdoors from \glspl{cnn}, discussed above.
We developed a defense approach specifically for \gls{rf} classifiers.
The concept behind our pruning defense is that leaves that are never used by samples in the validation set might be used by backdoored inputs. If these ``useless'' leaves are removed, performance of the classifier on the validation set should not decrease
%by a lot
significantly,
while decisions that are used for the backdoor are likely to be removed.
We developed several variants of our pruning defense with an increasing level of sophistication:
\begin{enumerate}[wide, labelwidth=!, labelindent=0pt]
\item Pruning leaves based on their usage: This means that the least used leaf is pruned first and the most used one last.
\item Like (1) but considering only ``benign'' leaves: We assume that attackers want malicious samples appear benign.
\item Like (2) but additionally using depth to decide when
%two % AH could be more than two
leaves are used equally often: The rationale is that we assume that hand-crafted backdoors require fewer rules to classify than regular samples and thus have lower depth.
\end{enumerate}

\autoref{fig:pruningExample} shows an example of variant (3) of pruning applied to a tree
\begin{figure}[b]
\includegraphics[width=\columnwidth]{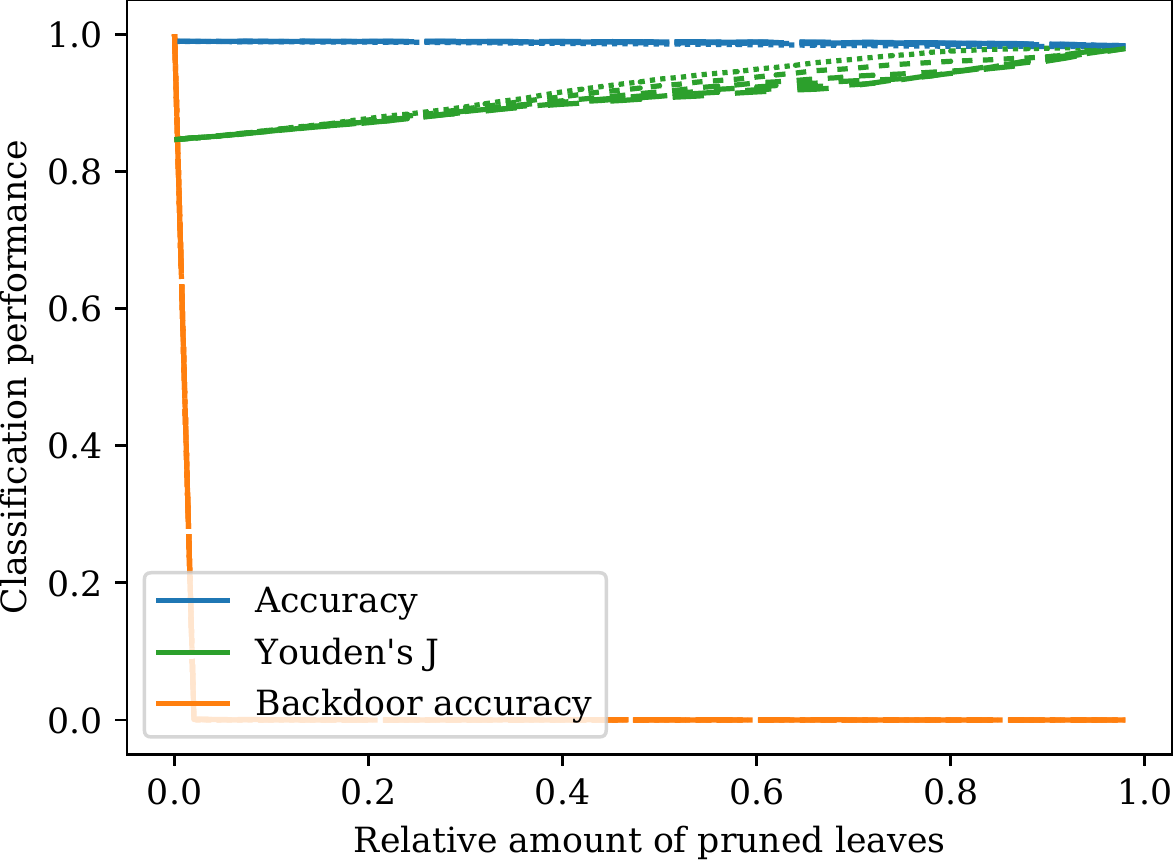}
\caption{Pruning an \gls{rf} for \unsw{}. The smaller the dashes, the smaller the validation dataset. The validation set size ranges from 1\% to 100\% of its original size.}
\label{fig:performancePruningRF}
\end{figure}
%\todo{Max: Which pruning algorithm actually?}
and \autoref{fig:performancePruningRF} shows the resulting accuracy for pruning according to variant (3) for the \unsw{} dataset. With sufficient leaves being pruned, the accuracy of the backdoor approaches zero. The accuracy for the regular data does not decrease significantly. It might even increase by reducing overfitting. With 90\% of leaves pruned, accuracy is still well above 99\%. Even with a very small validation set of just 1\% of its original size, the pruning still works as expected.

For \cic{} we get similar results, but pruning does not remove the backdoor completely ($\sim$30\% backdoor accuracy remains). We attribute this to the fact that in this dataset there are also regular flows which have $\text{stdev}(\text{TTL}) > 0$. Thus, backdoored flows cannot always be sharply distinguished from regular flows. We find that pruning only benign leaves is very beneficial for \unsw{} but not for \cic{}. However, considering depth when making the decision which leaf to prune first, leads to the backdoor being removed considerably earlier in the pruning process.

\section{Discussion}
From our experiments, we can make three main recommendations for the deployment of \gls{ml} models which have been obtained from a third-party.

To ensure that no backdoor is contained in the model,
%in question
it has to be  analyzed carefully for questionable decisions and potentially unnecessary features. For this purpose, \glspl{pdp} and \gls{ale} plots are an effective tool. In fact, already throughout the training process explainability plots constitute a useful tool to ensure that the model is not unintentionally trained  to artifacts the dataset yields.
On the other hand, the implementation of a backdoor as conducted in this research is only possible when using several features involving the \gls{ttl} value. Even though it might seem tempting to provide all possible features to a \gls{dl} or \gls{rf} classifier and let it learn the most important ones, this strategy should be avoided.

For \gls{rf} classifiers, which can be considered one of the most important classifiers for \glspl{ids}, the pruning technique we proposed is able to reduce backdoor efficacy significantly. At the same time, the classifier's detection performance is not substantially reduced. Thus, we recommend always including a validation set when providing a \gls{dt} or \gls{rf} to another party. Even if the validation set is significantly smaller than the training set, the defensive properties are still upheld.

%We found it
It is surprising that the neural network pruning and fine-tuning methods were ineffective for removing the backdoor from our \gls{dl} model in all our experiments. Since \glspl{cnn} are conceptually very similar to \glspl{mlp} it is not obvious that methods working for the former do not work for the latter. The difference that a \gls{cnn} always only looks at a portion of the input \cite{wikipedia_convolutional_2019} and not at all of it (unlike an \gls{mlp}) should not change the efficacy of the pruning approach. This leads us to the conclusion that probably the proposed methods are insufficient for \glspl{mlp} and more research is required to develop methods suitable for them.

\section*{Acknowledgements}
The Titan Xp used for this research was donated by the NVIDIA Corporation. We thank Prof.~Andreas Rauber for his suggestions to improve the explainability plots.

\bibliographystyle{ACM-Reference-Format}
\bibliography{bibliography}

\end{document}